\documentclass{ragtime}
\usepackage{graphicx} 
\usepackage{natbib}
\usepackage{aas_macros}

\newcommand{\mbh}{M_{\bullet}}
\newcommand{\msun}{M_{\odot}}
\newcommand{\mast}{M_{\star}}

\newcommand{\rg}{R_{\mathrm{g}}}

\newcommand{\rdd}{R_{\mathrm{d}}}
\newcommand{\rt}{R_{\mathrm{t}}}

\newcommand{\rast}{R_{\star}}
\newcommand{\rsun}{R_{\odot}}

\newcommand{\beq}{\begin{equation}}
\newcommand{\eeq}{\end{equation}}
\newcommand{\rK}{{\mbox{\tiny K}}}
\newcommand{\rd}{{\rm d}}

\title{A secondary orbiter under collisions with an accretion disk\footnote{Proceedings of RAGtime \#26--28 (2024/2025/2026): Workshops on black holes and neutron stars (Silesian University in Opava), in prep.}}
\author{Vladim\'{\i}r Karas}

\date{1$^{\rm st}$ January 2026}

\begin{document}
\maketitle

\begin{abstract}
Dynamics of stellar orbits in dense stellar systems and nuclear star clusters (NSC) with an embedded supermassive black hole (SMBH) is governed a complex interplay of different forces. In particular, star--star gravitational collisions (relaxation), physical collisions between stars, and the hydrodynamical interaction with any surrounding gaseous environment, such as an accretion disk. These processes influence the stellar distribution, the feeding of the central black hole, and the generation of observable phenomena. Furthermore, the self-gravity of the accretion medium modulates the long-term evolution, adding significant complexity to the system's dynamics. By employing elementary arguments we outline the mentioned influences in their mutual competition.
\end{abstract}

\section*{Star--disk interactions: gravitational and hydrodynamical effects in competition}

Let us turn our attention to the mutual interplay between two competing processes that are both capable of shaping dense stellar systems embedded into accretion flows in galactic centres \citep{1983ApJ...273...99O,1991MNRAS.250..505S}: (i)~mutual collisions between stars and gas
in a toroidal configuration or a quasi-spherical inflow (potentially with an emanating jet) versus (ii)~gravitational relaxation and direct collisions due to close encounters between individual stars of the cluster. 

Hydrodynamical effects encompass a range of interactions between stars and a gaseous environment. Stars passing through an accretion disk experience drag forces that dissipate orbital energy and momentum \citep{2000ApJ...536..663N}. This drag is primarily due to the gravitational interaction with the wake formed in the disk (dynamical friction in a fluid) and, depending on the system parameters but usually to a lesser extent, accretion of disk material onto the star. If a star is massive enough and embedded within the disk, it can create a gap in the disk and migrate inward, tidally coupled with the disk material. Otherwise, the excitation of density waves dominates, also leading to radial migration in either direction. The presence of embedded stars influences the chemical properties of the accreting medium \citep{1993ApJ...409..592A,1994ApJ...423..581A}. In case of inclined trajectories, high-velocity impacts with a dense disk can strip or ablate the outer layers of a star's envelope. The disk itself, even if its mass is a small fraction of the SMBH, exerts a non-spherical gravitational influence on stellar orbits, leading to oscillations in eccentricity and inclination, particularly through mechanisms like the Kozai-Lidov effect \citep{1962AJ.....67..591K,1962P&SS....9..719L}.

Gravitational relaxation describes the cumulative effect of small-angle encounters between individual stars. The energy and angular momentum of stellar orbits evolves on the two-body relaxation time over which the orbital properties are randomized \citep{1987degc.book.....S}. In a stellar system with a central SMBH, gravitational relaxation drives the formation of a stellar cusp, where the star density typically follows an approximately power law over the black hole's sphere of influence \citep{1976ApJ...209..214B}. For a typical SMBH, this timescale can be very long, of the order of a gigayear. 
The classical Bahcall-Wolf cusp model, derived from Fokker-Planck studies, simplifies assumptions, such as an isotropic stellar distribution, the system in steady state, and neglecting physical collisions. At very small radii, near the SMBH, tidal disruptions and direct consumption of stars entering the loss cone can modify the cusp. In fact, the systematic nature of hydrodynamical drag caused by the accretion disk leads to gradual flattening of the stellar system towards the disk plane; in consequence a recognizable stellar ring can form.

We want to estimate relative importance of the two channels  
and their influence upon long-term evolution of stellar population within $\approx1\,$pc from the galactic centre. To this end we assume that a massive ($M\approx10^6_\odot$--$10^9_\odot)$ dark object, presumably a black hole, is located in the core and accretes gas from the surrounding disk. The orbiter can be interpreted either as one of the stars originating from the cluster ($M_{\star}\simeq M_{\odot}\ll M_{\bullet}$), or a companion intermediate-mass black hole {$M_{\star}\simeq 10^{3\mbox{--}5} M_{\odot} < M_{\bullet}$. The nature of the orbiter determines how efficient is its interaction with
the accretion flow (either a largely laminar flow around an obstacle versus a development of wake turbulence), what are the boundary conditions imposed on the flow at its inner boundary (hard surface of a compact body versus a free, supersonic infall onto a black hole) and far from the center (equatorial, planar inflow versus a twisted and inclined structure), and how this object interacts with other members of the nuclear cluster -- via dynamical friction, collective effects, and related. Also, accretion disk crossings are more likely to be significant in systems with high SMBH mass as compared to lower SMBH masses \citep[the relaxation time is longer for more massive black holes;][]{2020ApJ...889...94M}.

In the early papers, the two main ingredients of the model were often considered separately. This approach helps us to disentangle the role of different effects, however, it is not fully self-consistent and it may hide their mutual interaction. Concerning the direct star-disk collisions, \citet{1994ApJ...434...46Z} and \citet{1998MNRAS.298.1069I} discussed local physics of hydrodynamic collisions between the accretion disk and the orbiting star or a black-hole companion, while \citet{1991MNRAS.250..505S} and \citet{1998MNRAS.298...53V} examined the evolution of orbital parameters of individual satellites. \citet{2002A&A...387..804V} derived time-scales that govern the orbital evolution of perturbed satellites as they gradually sink towards the center.

In dense star clusters, particularly close to an SMBH, physical collisions between stars become a significant evolutionary factor \citep{1999ApJ...514..725R,2001A&A...375..711F,2002A&A...394..345F}. These collisions can lead to various outcomes, including a complete disruption, mass loss, or mergers. At high relative velocities, typical near an SMBH, collisions can be disruptive rather than leading to mergers. Stellar collisions are highly effective at refilling the loss cone, often dominating over relaxation-induced refilling for tidal disruptions in dense regions. Collisions can thus dominate the dynamical evolution over two-body relaxation and tidal effects within the inner parts of the black hole's sphere of influence, particularly in very dense nuclei.

Low-level AGN accretion disks or rarefied gaseous environments (like in the Galactic center's Sgr A*) are generally not dense enough for hydrodynamic effects to dominate orbital histories of compact objects. For stellar-mass black holes and neutron stars, the gravitational cross-section is very small, and the changes to orbital properties from disk interaction are typically negligible compared to radiation reaction or stellar relaxation. Non-compact stars (main-sequence stars, giants) have larger geometric cross-sections, making them much more susceptible to hydrodynamic drag than compact objects. Giants, with their extended atmospheres, are particularly vulnerable to mass loss and removal from the inner regions due to hydrodynamical interaction.

The course of gravitational relaxation has been approached under different approximations. We are mainly interested in systems that
have already achieved their quasi-equilibrium stage, and hence we 
can concentrate on collisional relaxation between the members of the cluster rather than (much faster) process of violent relaxation, Landau damping and phase mixing which appear to be dominant during the initial stages \citep{1991ApJ...375..687M}. Nevertheless, we have to include losses of energy by the satellites due to gravitational friction in the mean field
of the entire cluster  \citep{1987MNRAS.224..349B,1998ApJ...502..167C}. More recently, the great progress has been achieved by numerical simulations of the $N$-body problem which employs the special-purpose techniques and hardware; see e.g. \citet{1997ApJ...478...58M}.

\begin{figure}[tbh!]
\center
\includegraphics[width=\textwidth]{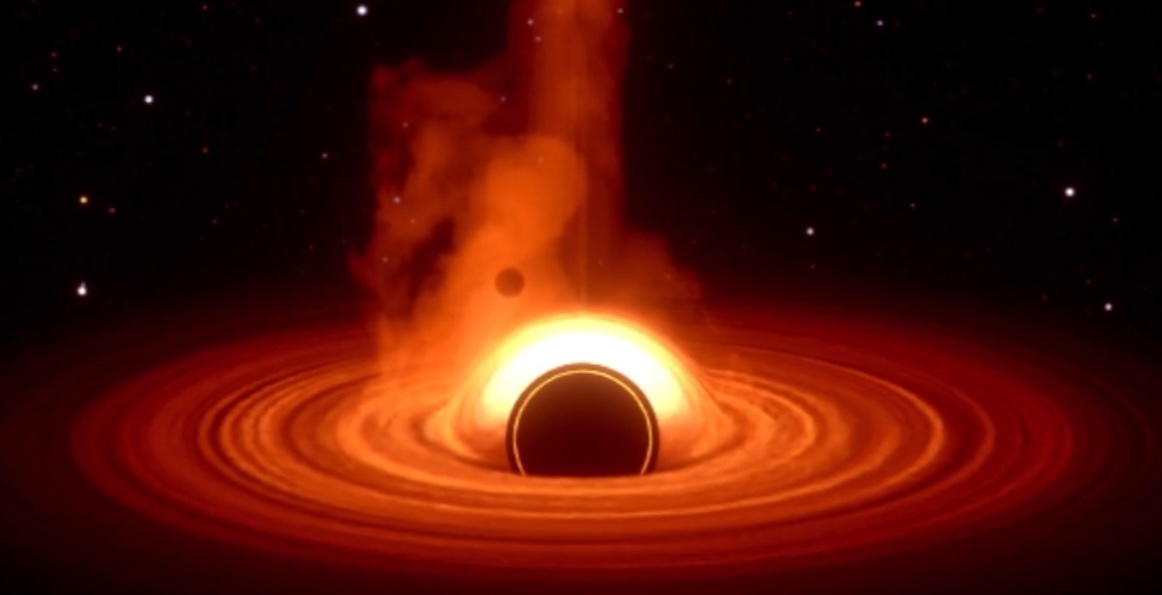}
\caption{A supermassive black hole is surrounded by a gaseous environment of an accretion disk, possibly also a jet/wind outflow, and a less-massive secondary orbiter that periodically perturbs the signal in electromagnetic as well as gravitational domains \citep{2021ApJ...917...43S,2023ApJ...957...34L,2024SciA...10J8898P}.}
\label{fig1}
\end{figure}

It has been recognized \citep{1994MNRAS.267..557P,1995MNRAS.275..628R} 
that successive star-disk collisions with a gaseous environment (of astrophysically realistic density values and profile) significantly influence the distribution function of the stars in the inner regions: one cannot neglect the dissipative collisions. On the other hand, the relaxation processes should not be ignored when collisions cease to operate, which is likely to occur on very long time-scales and within a sufficiently dense cluster (we adopt central densities of the order $\approx10^6$--$10^8M_\odot\,$pc$^{-3}$ in the cluster). While gravitational relaxation drives the formation of spherically symmetric cusps, hydrodynamical interactions can lead to the formation of flattened, anisotropic stellar distributions or even toroidal structures in the inner parts of the cluster, where stars are dragged into the accretion disk plane \citep{1991MNRAS.250..505S,2001A&A...376..686K}.

In summary, the main objective here is to estimate mutual importance of the mentioned collisions versus relaxation. Clearly we need to introduce various simplifications in the treatment. Then we will be able formulate equations and solve them numerically to find regions in the parameter space where star-disk collisions dominate over relaxation, or vice versa it is the gravitational interaction that governs the system evolution. Let us also note that the interaction of stars with the gaseous environment can contribute to the fueling of the central black hole and its overall activity reaching beyond the rates provided by two-body relaxation alone.


The gravitational relaxation primarily shapes the large-scale, spherical stellar distribution over long timescales, leading to a stellar cusp. However, hydrodynamical effects become increasingly dominant for stars on highly eccentric orbits passing through dense accretion environments and leading to significant changes in orbital elements, potential stellar destruction, and flattening or re-shaping of the inner stellar cluster on relatively short timescales. The interplay between these mechanisms is crucial for understanding the complex dynamics in galactic nuclei. The gravitational waves emission takes over in the late stages of the orbital evolution preceding the final plunge or tidal disruption of the orbiter.

\subsection*{Simple approach to star-disk hydrodynamical collisions} 

The cumulative effect of many passages through the disk is often referred to as ``the grinder''; it systematically modifies the star's orbit:
(i) the orbital decay leads to shrinking the semi-major, thus the change of orbital energy and angular momentum. Under the standard set-up of the accretion disk parameters, stars gradually {\em lose} energy and angular momentum, pulling them inward to the central black hole. This process is associated with (ii) the orbital circularization, where the fractional change in orbital energy is typically larger than the fractional change in angular momentum. This generally results in a monotonic decrease of eccentricity. The drag forces lead to (iii) the alignment of stellar orbits with the plane of the accretion disk, causing the inclination to decrease and eventually embedding the orbiter in the disk plane.

In the first stage, let us assume the accretion disk to be geometrically thin
\citep[a.k.a. standard accretion mode]{1973A&A....24..337S,1974ApJ...191..499P}. The accretion disk rotates with Keplerian velocity $v_\rK=\sqrt{GM/r}$. This approximation enables us to characterize it by $z$-averaged quantities. The surface density of an axially symmetric disk depends only upon the radial distance from the centre,
\beq
\Sigma_\rd=Kr^s\Sigma_\ast\, ,
\eeq
where $\Sigma_\ast\equiv m_\ast/(\pi R_\ast^2)$ is the surface density associated with the star. $K$ and $s$ are free parameters that can be derived from the Shakura-Sunyaev approximation. 

The star--disk interaction is described as a nonelastic collision of the star and the disk material that lies along the star's
trajectory; this is substantiated by the supersonic nature of the motion \citep{1994ApJ...434...46Z}.
Provided $\bold{v}_\ast=(v_r,v_\vartheta,v_\varphi)$ and
$\bold{v}_\rd=(0,0,v_\rK)$ are the star and the accretion disk velocity before the interaction, the velocity $\bold{v}'$ of both the star and the expelled disk material can be evaluated 
through the momentum conservation law:
\beq
\mathbf{v}'={\frac{1}{A+1}} \left( v_r\,, v_\vartheta\,, 
v_\varphi + Av_\rK\right)\,,
\label{eq:v'}
\eeq
where
\beq
A\equiv{\Sigma_{\rd}\over\Sigma_\ast}{v\over v_\vartheta}=
{\pi R^2_\ast\Sigma_{\rd}\over m_\ast}{v\over v_\vartheta}\,.
\label{eq:A}
\eeq

Requirement of supersonic velocity poses limits upon
the orbital parameters. We can start with a rough estimate of a lower limit of inclination: Let's assume the star 
moves on a circular trajectory with Keplerian velocity $v_\rK= \sqrt{ 
GM/r} =2.12 \times 10^{10} (r/r_g)^{-1/2} \rm cm\, s^{-1}$ 
and inclination $i$. Magnitude of the relative velocity is then
\beq
v_{\rm rel} = 2v_\rK\sin(i/2) = 4.24 \times 10^{10} \sin(i/2)
\left({r \over r_g} \right)^{-1/2}\, \rm cm\,s^{-1}\,.
\eeq
Speed of sound in the standard accretion disk for
the central mass $M_{\bullet}=10^8M_\odot$, accretion rate $\dot{M} = 0.01 \dot{M}_{\rm E}$, and the viscosity parameter $\alpha=0.1$ is
\beq
v_{\rm s} = 1.27 \times 10^7 \left({r \over r_g}\right)^{3/8} \,
\rm cm\,s^{-1}\,.
\eeq
Condition $v_{\rm rel} \gg v_{\rm s}$ then reads:
\beq
\sin(i/2) \gg 3.1 \times 10^{-4} \left({r \over r_g}\right)^{7/8}\,.
\eeq

In the region where the central object dominates the gravitational field and the gaseous environment is confined to a geometrically thin, perfectly planar accretion disk, 
stars move between subsequent passages along (almost) unperturbed
orbits (Keplerian ellipses in the Newtonian regime).
These are characterized by the semi-major axis $a$, eccentricity $e$, inclination $i$ with respect to the plane of the disk, and the longitude $\psi$ of apocentre, measured from one of the nodes. Equivalently, the orbit can be
characterized by a set of variables $a$, $x=\cos i$,
$y=1-e^2$ and $z=1-e\cos\psi$.
Through the disk, we can evaluate changes of the star's binding
energy $E$, angular momentum $L$ and its $z$-component $L_z$ 
and the radial component of velocity during each individual passage \citep{1999A&A...352..452S},
\begin{eqnarray}
\delta E = -\delta v^2 & = & {A\over ay} \left[ 2z -y
+ xz^{3/2}\right] \label{eq:delta1}\\
\delta L = r\delta v_{\rm T} &=& A \sqrt{ ay \over z}
\left[ x - \sqrt{z} \right] \label{eq:delta2}\\
\delta L_z = r\delta v_\varphi &=& A \sqrt{ ay \over z}
\left[ 1 - x\sqrt{z} \right] \label{eq:delta3}\\
\delta v_r &=& A \sqrt{ 2z - z^2 -y \over ay}
\label{eq:delta4}
\end{eqnarray}
As $E=1/2a$, $L=\sqrt{ay}$, $L_z=\sqrt{ay} x$
and $v_r=-\sqrt{(2z-z^2-y)/ay}$, we obtain
\begin{eqnarray}
\delta E &=& -{1\over 2a^2} \delta a\label{eq:delta5}\\
\delta L &=& {1\over 2}\sqrt{y \over a} \delta a +
{1\over 2}\sqrt{a \over y} \delta y \label{eq:delta6}\\
\delta L_z &=& \delta L + \sqrt{ay} \delta x
\label{eq:delta7}\\
\delta v_r &=& {1 \over 2\sqrt{ (2z - z^2 -y) (ay)}} \delta y -
{1-z \over \sqrt{ (2z - z^2 -y) (ay)}} \delta z \nonumber \\
& & + {\sqrt{ 2z - z^2 -y} \over ay} \delta L \label{eq:delta8}
\end{eqnarray}
Substituting (\ref{eq:delta5}) $\sim$ (\ref{eq:delta8}) into
(\ref{eq:delta1}) $\sim$ (\ref{eq:delta4}) and rewriting differences $\delta$ into differentials,
\begin{eqnarray}
\rdd a &=& A {2a \over y} (y - 2z + xz^{3/2}) \label{eq:dif1}\\
\rdd x &=& A {1 - x^2 \over \sqrt{z}} \label{eq:dif2}\\
\rdd y &=& 2A \left( 2z - 2y + {xy \over \sqrt{z}} - xz^{3/2}
\right) \label{eq:dif3}\\
\rdd z &=& 2A (x \sqrt{z} - 1) \label{eq:dif4}
\end{eqnarray}
In the adopted approximation the solution for the orbit shape
$[a(x),y(x),z(x)]$ is independent of the disk density profile (incorporated in the variable $A$), although the temporal evolution is influenced.

The characteristic grinding time for an orbit to be dragged into the plane of an accretion disk is typically of the order of $10^4$ to $10^7$ orbital periods, or $10^4$ to $10^7$ years, depending on stellar type and disk surface density \citep{1991MNRAS.250..505S,1998MNRAS.293L...1V}.
The highly eccentric orbits experience the most significant effects from disk interactions. Retrograde orbits that are 
initially counter-rotating with respect to the disk experience a higher relative velocity with the disk gas, thus 
leading to a faster evolution (accelerated orbital decay) compared to prograde orbits. While the general trend is orbital decay, circularization, and alignment, other factors introduce significant complexities. In particular, non-spherical gravitational potential of a massive disk or a third massive body can induce long-term oscillations in a star's eccentricity and inclination \citep{1998MNRAS.298...53V,1999A&A...352..452S,2016ApJ...822...25H}. These secular oscillations can pump the osculating eccentricity to very high values, which in turn accelerates the orbital decay as the star passes through the disk at high velocity, making the hydrodynamical drag more efficient \citep{2007A&A...470...11K}. This mechanism is particularly relevant for intermediate-mass black holes, as the damping effects (like relativistic pericenter advance or the cluster mean field effect) become more dominant for higher SMBH masses. The presence of the disk's gravitational field means that the Kozai--Lidov mechanism {\em must} be considered alongside 
dissipative hydrodynamical drag.

Stars, especially the bloated red giants, can have their outer envelopes stripped or ablated by the ram pressure of disk-crossing impacts, leading to the mass loss. This process is particularly efficient for low-surface-density stars \citep{1996ApJ...470..237A}. The mass loss contributes to fueling the SMBH \citep{2020ApJ...889...94M}. Self-gravitating disks are then prone to fragmentation and formation of a clumpy structure. The clumps can significantly affect star-disk interactions, potentially enhancing the role of mass stripping of red giants \citep{2020ApJ...903..140Z}. On the other hand, if a star is sufficiently massive relative to the disk's properties, it can clear an annular gap in the disk (analogy with gaps in proto-planetary disks). This prevents further accretion onto the star. Once a gap is formed, the star's motion becomes tidally coupled to the disk and it can migrate radially (inward or outward) on the disk's viscous evolution timescale \citep{1993prpl.conf..749L}. The direction and timescale of migration depend on the star's location relative to the disk's mass distribution.

Ablation of stars by relativistic jets emanating from the central supermassive black hole (SMBH) acts on stars that pass near the jet axis (presumed to coincide with the rotation axis of the central black hole). It has quite different impact on stellar orbits and the overall cluster structure: star-disk interactions involve a gaseous accretion disk, while star-jet interactions involve a collimated, relativistic plasma outflow. The primary mechanism is ram pressure exerted by the jet on the star's outer layers \citep{2017bhns.work....1A,2020ApJ...903..140Z,2025MNRAS.540.1586K}. After jet encounters, stars can be collisionally heated. Their subsequent enlargement in size can further enhance mass removal during passages if the star does not cool down quickly (classified as {\em warm colliders}). Ablation changes the stellar appearance, making the stars apparently warmer, bluer and fainter in near-infrared bands compared to what would be expected from a normal evolution. Dynamical processes such as resonant relaxation (RR) can also change the inclination of stellar orbits, increasing the number of stars that cross the jet path over the jet's lifetime \citep{1996NewA....1..149R,2016MNRAS.458.4143S}. 

Jet precession enlarges the volume affected by the jet, thus influencing more stars.  Bow shocks are formed during star-jet interactions and they can act as sites of particle acceleration, contributing to the production of high-energy, namely, gamma-ray emission and neutrinos from AGN.

The mass lost from stars due to star-disk interactions can potentially fuel the SMBH. In contrast, the material lost from stars due to jet ablation is often ejected outward along the jet direction, contributing to jet mass-loading and outflow phenomena \citep{2021ApJ...917...43S}. The efficiency of these mechanisms varies with the distance from SMBH. For instance, in the Galactic Center, tidal stripping is efficient at very small radii ($<1$ mpc), jet-induced ablation is most profound for red giants in the S-cluster (about 0.006--0.04 pc, and up to $\sim 0.1$ pc), while star-clumpy disk collisions are expected to be efficient at larger radii (about 0.04--0.5 pc), particularly in regions where the disk is gravitationally unstable.

Stars on eccentric orbits are prone to tidal disruption (or partial tidal disruption) by the central massive black hole at the moment of pericenter passage. 
The process is driven by low angular momentum trajectories that have their periapses within the black hole tidal radius,
\begin{equation}
\rt=\left(\frac{\mbh}{\mast}\right)^{1/3} \!\rast
\simeq 10^{3} M_4^{-2/3}
\left(\frac{\mast}{\msun}\right)^{-1/3}
\left(\frac{\rast}{\rsun}\right) \,\rg,
\label{eq:Rt}
\end{equation}
where $\mast$ and $\rast$ are the stellar mass and radius. The process of tidal disruption introduces the notion of loss cone, which becomes emptied on a time scale of $\simeq{P}$, provided the perturbing processes are not strong enough to deflect stars away from their plunging orbits.

By increasing eccentricity to high values, Kozai oscillations can cause the pericenter distance of stellar orbits to shrink, thus effectively enlarging the loss cone of stellar orbits that are susceptible to tidal disruption. This enhances the rate of tidal disruption events that deplete highly eccentric orbits from the nuclear cluster's phase space \citep{1999MNRAS.303..483S,2020ApJ...889...94M}. Note that for black holes accreting near the Eddington limit, star-disk interactions define a ``disk loss cone'' in the orbital phase space. This loss cone is larger than the stellar tidal disruption loss cone, meaning that highly eccentric orbits are depleted by the disk interactions before they can reach the tidal disruption radius. However, stars can still move over the disk loss cone into the tidal disruption loss cone via large-angle scatterings.

The Kozai-type effects can be suppressed by other perturbations: (i) Strong gravity near the SMBH causes relativistic precession of stellar orbits, which can inhibit Kozai oscillations; (ii) The gravitational potential of the extended stellar cluster itself can cause secular precession and thus attenuation of Kozai oscillations. The overall impact on tidal disruption rates depends on the competition between these effects, with the enhancement being more pronounced for intermediate-mass black holes. 

The dispersed material provides a natural source to feed the accretion process by replenishing the inner disk. This process has been shown to operate in the system of a massive black hole -- accretion disk -- nuclear stellar cluster, nonetheless, its efficiency is affected by the pericentre precession that is driven by a combination of relativistic effect and Newtonian influence of the outer cluster, and a self-gravitating accretion disk or a torus \citep{2007A&A...470...11K}. This enhances the rate of stars plunging close to the central black hole, thus triggering the episodic supply of material onto the black hole. The accretion rate is then raised and the activity of the system is enhanced on the characteristic time scale, typically $T_\mathrm{K}\simeq10^5$--$10^6$~yrs for on which the loss cone becomes depleted. This feeding mechanism is more efficient for less-massive (tentative) intermediate mass black holes compared to the supermassive ones.

For compact objects (black holes and neutron stars), the hydrodynamic drag in the inner regions becomes generally negligible compared to gravitational radiation at the late stages of inspiral. However, it can produce a distinctive signature by always decreasing orbital inclination (forcing orbits towards equatorial prograde configuration), unlike radiation reaction which can increases it. This could be detectable with gravitational wave observatories \citep{2008PhRvD..77j4027B}. 
Star-disk interactions can even modify the mass function of the inner cluster, causing it to depart from its initial form. Light and/or compact stars migrate relatively slowly, leading to changes in their abundance relative to heavier stars. Finally, self-gravitating accretion disks can be sites of significant star formation with the newly formed stars then interacting with the accretion flow \citep{1999A&A...344..433C,2004CQGra..21R...1K}.

\subsection*{Summary}
While two-body relaxation drives the large-scale evolution and cusp formation in stellar clusters, stellar collisions dominate the innermost regions, particularly in dense nuclei, by causing disruptions and refilling the loss cone. Star-disk collisions are highly effective at refilling the loss cone for tidal disruptions, often dominating over relaxation-induced refilling in dense regions. Overlaid on these processes, the gaseous environment of an accretion disk introduces powerful dissipative forces that shrink and circularize orbits, but also non-dissipative gravitational forces (like the Kozai mechanism) that can dramatically increase eccentricity. The complex interplay of these mechanisms shapes the distribution of stars, fuels the central black hole and influences the observable properties of galactic nuclei. 

The idea of mutual interaction between stars and the environment of an accretion disk have gained new impetus in the context of  long-term evolution of the orbit of a satellite (intermediate-mass) black hole captured by a massive galactic nucleus. Repetitive transits across the accretion disk slab create a potent turbulent wake that needs to be further studied in order to understand the impact of dragging and the amount of material pushed out of the disk plane. The hydrodynamical drag competes with the orbital decay by gravitational waves. Renewed interest in the long-term evolution of the orbits under the influence of interaction with an accretion disk has emerged in the context of Repetitive Nuclear Transients and Extreme/Intermediate-Mass-Ratio Inspirals near SMBH \citep[e.g.,][and further references cited therein]{2024ApJ...973..101L,2025MNRAS.540.1586K,2025ApJ...978...91Y}.

\subsection*{Acknowledgement}
The author acknowledges fruitful discussions and long-term collaboration on related topics with drs.\ Václav Pavlík, Petra Suková and Michal Zajaček. Filip Houdek was helping with an artistic rendering in \textsf{Blender} for Fig.~\ref{fig1}. The exchange mobility project between the Astronomical Institute of the Czech Academy of Sciences and the Center for Theoretical Physics of the Polish Academy of Sciences, titled ``Appearance and dynamics of accretion onto black holes'', is greatly appreciated. 

\bibliography{references}

\end{document}